\documentclass[12pt, titlepage]{aastex62}
\pdfoutput=1
\NewPageAfterKeywords
\newcommand{\dedalus}{\href{http://dedalus-project.org}{Dedalus}}
\hypersetup{
  colorlinks,
  linkcolor=red,
  urlcolor=olive,
  citecolor=gray
}

\begin{document}

\title{Perspectives on Reproducibility and Sustainability of Open-Source Scientific Software from Seven Years of the Dedalus Project\\{\footnotesize A white paper submitted to the National Academies of Sciences, Engineering, and Medicine's Best Practices for a Future Open Code Policy for NASA Space Science}}
\collaboration{The Dedalus Collective}
\author{Jeffrey S. Oishi}
\affiliation{Bates College}

\correspondingauthor{Jeffrey S. Oishi}
\email{joishi@bates.edu}

\author{Benjamin P. Brown}
\affiliation{University of Colorado, Boulder}
\author{Keaton J. Burns}
\affiliation{MIT}
\author{Daniel Lecoanet}
\affiliation{Princeton}
\author{Geoffrey M. Vasil}
\affiliation{University of Sydney}

\section{Introduction}
\label{sec:intro}
As the Science Mission Directorate contemplates establishing an open code policy, we consider it timely to share our experiences as the developers of the open-source partial differential equation solver \dedalus{}. \dedalus{} is a flexible framework for solving partial differential equations. Its development team primarily uses it for studying stellar and planetary astrophysics.  \dedalus{} was developed originally for astrophysical fluid dynamics (AFD), though it has found a much broader user base, including applied mathematicians, plasma physicists, and oceanographers. Here, we will focus on issues related to open-source software from the perspective of AFD. We use the term AFD with the understanding that astrophysics simulations are inherently multi-physics: fluid dynamics coupled with some combination of gravitational dynamics, radiation transfer, relativity, and magnetic fields. In practice, a few well-known open-source simulation packages represent a large fraction of published work in the field. However, we will argue that an open-code policy should encompass not just these large simulation codes, but also the input files and analysis scripts. \textbf{It is our interest that NASA adopt an open-code policy because without it, reproducibility in computational science is needlessly hampered.} 

With the caveat that we have not performed bibliometric analyses, it seems reasonable to assert that the majority of papers published in AFD are done using open-source codes, especially in star formation, accretion disk physics, and large-scale structure formation. \emph{Athena}, \emph{Enzo}, \emph{Pencil Code}, \emph{Pluto}, and \emph{Gadget 2} are all widely used open-source tools. In stellar AFD, there are fewer examples of open-source tools; \href{http://magic-sph.github.io}{MagIC} and \href{http://amrex-astro.github.io/MAESTRO/}{MAESTRO} are important exceptions. For many of these simulation codes, the preferred method of in-depth data analysis is via \href{http://yt-project.org}{yt}, a pioneering open-source project that introduced many of the important development and community-building concepts we use in \dedalus{}. These codes tend to be very large: Athena has $\simeq 80,000$ SLOC; Enzo is over $120,000$ lines, and MAESTRO is over $161,000$ lines\footnote{All source lines of code reported here are measured using the the \href{https://github.com/bytbox/sloc}{sloc tool} on publicly available repositories of each code as of 3 January 2018}! Such complexity is not unique to open-source tools; two of us have significant experience with closed-source astrophysics codes (\emph{ORION} and \emph{ASH}) of comparable size. The sheer size of these codes presents challenges to reproducibility: it takes a significant amount of effort to even be able to build and run them. 

The sub-community of astrophysical fluid dynamics modeling thus tends to embrace open-source even without an official policy. However, this is, in our view, not sufficient, especially as it can vary from field to field: many of the most important spherical dynamo codes for studying stellar and planetary magnetic fields remain closed-source (e.g. \emph{Rayleigh}\footnote{At the time this paper was submitted on 12 January 2018, \emph{Rayleigh} was closed source. Since that time, it has been made available under the GPL at \url{https://github.com/geodynamics/Rayleigh}}, \emph{ASH}, various derivatives of the Glatzmaier/Gilman code). The simulation code is only one part of a long pipeline that starts with an idea and ends with a published result. An open-code mandate should eliminate any closed-source bottlenecks that could hamper what is, in our view, the cornerstone of the scientific computing enterprise: repeatability. That is, simply because the largest piece of software is available, if the input files and analysis scripts are not also public, reproducibility becomes impossible. 

\dedalus{} is written primarily in python; it has a very small code base of approximately 7000 source lines of code (SLOC). It originated in an effort by one of us to develop a easily modifiable, modern fluid dynamics solver for studying magnetohydrodynamic turbulence problems in AFD. The first commit to the project was in 2010. In 2012, the project expanded from two people to five. Since then, nine people have made commits to the code base, including four not affiliated with the core development team. We chose the term ``core team'' to represent not a privileged set of developers but a reflection of the current number of highly active participants in the development and future of \dedalus{}. We have logged 46 pull requests, representing peer-reviewed code contributions. The code is licensed under the GNU General Public License Version 3 (GPL). We chose this license to ensure that the source code always be available for inspection and auditing by any interested parties. 

We do not ask users to register before receiving the code. Doing so would, in our view, constitute a barrier to access without providing any substantive estimate on the number of users. Instead, we can estimate the size of our community by a number of metrics. We have 177 unique users participating in the \texttt{dedalus-users} mailing list for discussions of using the code, and 13 unique users participating in the \texttt{dedalus-dev} for discussions of development of new features. \dedalus{} has been used in at least thirteen peer-reviewed publications \citep{2018Bordwell,2017PhRvF...2h3501A,PhysRevFluids.2.094804,2017ApJ...841....1C,2017ApJ...841....2C,2017MNRAS.466.2181L,2016ApJ...832...71L,2016JCoPh.325...53V,2016PhRvE..94e3206D,2016PhRvL.116j5004D,2016MNRAS.455.4274L,2015PhRvE..91f3016L,2014ApJ...797...94L}.

Here, we argue that the benefits of open-source projects far exceeds their costs; we outline our view on exactly what those benefits and costs are, and we suggest that in order to maximize scientific returns on software development, \emph{open-source} is not enough. The best use comes from \emph{community-driven} software tools that blur lines between users and developers. We strongly recommend that Science Mission Directorate adapt an open code policy that embraces \emph{explicitly licensed}, discoverable software. 

\section{Reproducibility}
\label{sec:repro}

Reproducibility is the cornerstone of scientific research, but there are serious concerns regarding the ability to reproduce computational results \citep[see][and references therein]{2012ICERM_REPORT}. There are many definitions of reproducibility; here we focus on the ability of outside researchers to replicate published results. Open-source codes support this reproducibility by allowing any researcher the ability to, in theory, use the same tool as the original study. However, as mentioned in the introduction, modern AFD codes are extremely large; though \dedalus{} is an order of magnitude smaller than others, it gives users much more freedom in specifying problems and thus remains similarly complex to use. This complexity makes the cultivation of a \emph{community} of contributors important in ensuring the ability of outside researchers to reproduce results. 

We actively work to build a community of practice influenced by the \emph{Enzo} and \emph{yt} communities, as outlined in \citet{2013arXiv1301.7064T}. This gives researchers a way not only to access the source code but also ask questions and contribute to the code. However, community of this sort is still not enough to ensure reproducibility. In addition, several members of our community have made publicly accessible the version control repositories containing \emph{all} material required to reproduce the results published in \citet{2017ApJ...841....1C,2017ApJ...841....2C}\footnote{\url{https://bitbucket.org/jsoishi/weakly_nonlinear_mri}}, \citet{2017PhRvF...2h3501A}\footnote{\url{https://bitbucket.org/exoweather/polytrope}}, and \citet{2018Bordwell}\footnote{\url{https://bitbucket.org/exoweather/polytrope}}. By publishing analysis scripts and input files, these repositories allow outside researchers to download \dedalus{}, run it using the same inputs, and then analyze the output to produce the plots found in those papers. We believe that these are the first steps in a process that can improve the quality and reproducibility of the science results while offering an opportunity for outside researchers to build on those results. Researchers can test the claims made in each paper and then extend those claims to answer new questions that the authors may not have envisioned. For large simulations, the amount of computational resources required to \emph{directly} re-run all the simulations may be prohibitive. In this case, mechanisms to allow access to the output data from the simulations should be made available. We discuss infrastructure to do this in section~\ref{sec:future}.


In addition to reproducing published results--an action we suggest is far too rarely performed--open-source policies help promote verification and validation testing. In AFD, verification tests are plentiful; there is a strong tradition of benchmark problems and inter-code comparisons in AFD \citep[e.g.][among many others]{2001JGR...106.3715B,2014GeoJI.197..119M,2014ApJS..210...14K,2016MNRAS.455.4274L}. However, direct validation is nearly impossible for AFD: there are precious few astrophysical experiments that can be done repeatably in the lab. In lieu of this, AFD codes can perform validation tests against standard terrestrial fluid dynamics problems. Here again, \dedalus{} has been significantly aided by its community: by eliminating boundaries between users and developers, we actively encourage the development of validation problems which are then included in the code base (e.g. \href{https://bitbucket.org/dedalus-project/dedalus/src/cf57edf1516b931bc4ad0a7895d06219bbe90414/examples/ivp/3d_rayleigh_benard/rayleigh_benard.py?at=default&fileviewer=file-view-default}{Rayleigh-Benard Convection}). We have found that these validation problems are often the starting point for new users who use them as examples and extend them to their problem. Thus, validation problems also communicate best practices for using the tool in a way that has been very beneficial for the entire community. 

\textbf{We would suggest that any open code policy adopted by the SMD explicitly consider the importance of releasing input files, analysis scripts, and build-time parameters.} We elaborate on the need to consider the software stack upon which the simulation and analysis tools rely in section~\ref{sec:stack}.


\section{Maintenance and Support}
\label{sec:support}

Providing open-source code is not without cost. Direct costs are quite minimal; they are limited to web hosting. The source for \dedalus{} is hosted at \href{https://bitbucket.org}{Bitbucket}, a commercial provider of distributed version control services that allows open-source projects to be hosted for free. An often-discussed cost of open-source research products is the ``need'' to provide user support for them. However, we have found that, despite a community that spans many scientific disciplines, the amount of direct support we do is small and it occasionally leads to significant improvements to the code base itself. Indirect support, in the form of responding to bug reports and implementing enhancements suggested by community members is also small, amounting to no more than a few person hours per month. Moreover, by encouraging a community of practice in which there is not a clear distinction between ``users'' and ``developers'', we have mechanisms for anyone to provide fixes and receive credit in the form of having their identity embedded along with the changes in the version control log. 

\begin{figure}
  \centering
  \includegraphics[width=0.5\textwidth]{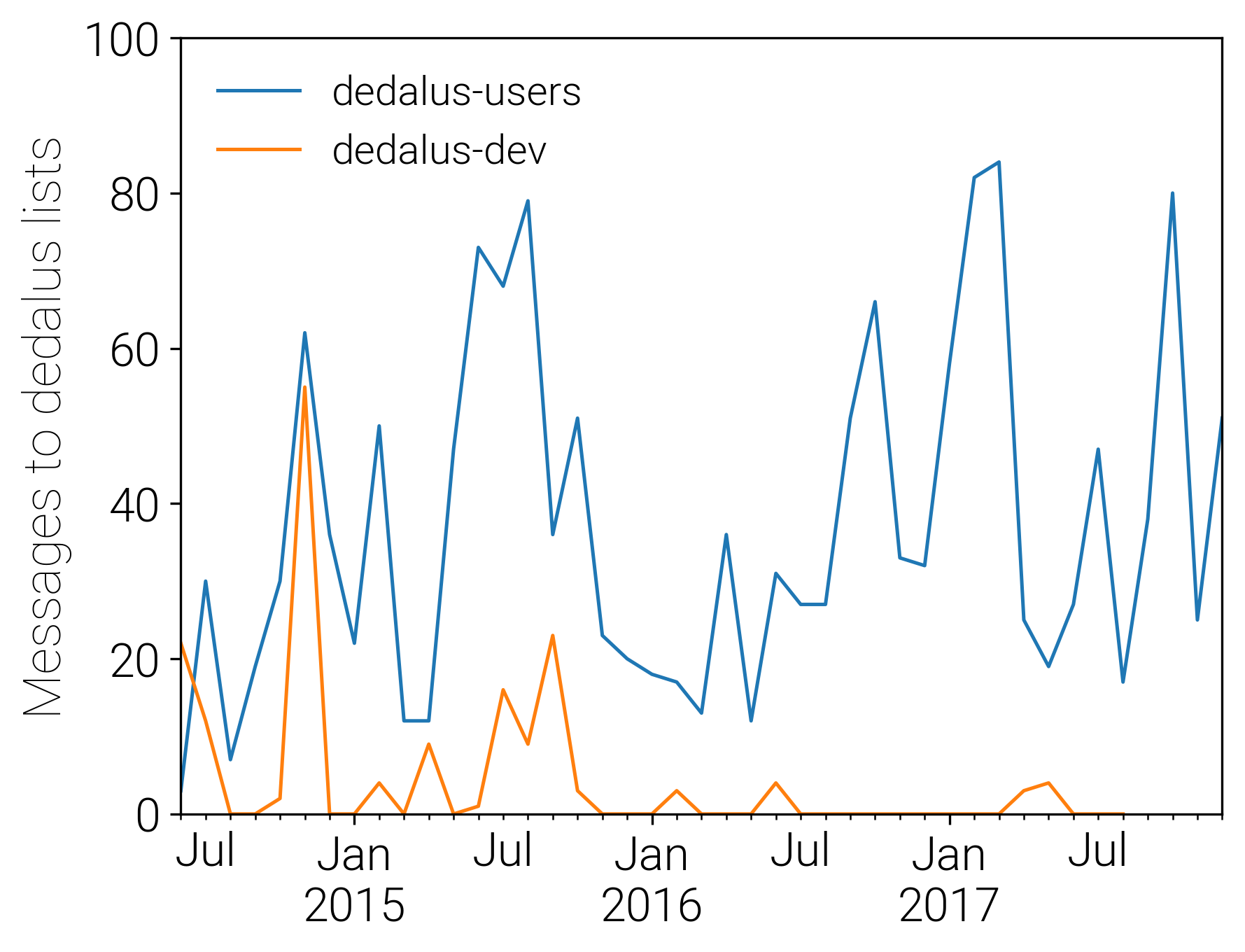}
  \caption{Message traffic on the \texttt{dedalus-users} and \texttt{dedalus-dev} google groups.}
  \label{fig:messages}
\end{figure}

Figure~\ref{fig:messages} shows the number of messages per month sent to the the two mailing lists we maintain, \texttt{dedalus-users} and \texttt{dedalus-dev}. \texttt{dedalus-users} shows a steady (roughly linear) average growth over the past two and a half years, with a peak of just over 80 messages in one month. Making a rough assessment of 15 minutes per message, this represents approximately 10 person-hours per month on the part of the core team, if we assume that half of the messages are questions and the other half responses. Given a core team of 5 researchers, this represents an upper limit of 2 hours per month per developer. However, non-core members of our community often \emph{answers} questions on this list in addition to asking them (and core team members also \emph{ask} questions too!). The groups themselves are archived and searchable, making them a knowledge base for users. Finally, we note that often questions on the user mailing list lead to the development of new example problems or features: recently, a question on \texttt{dedalus-users} about how to handle the arguments to a certain kind of function object\footnote{\url{https://groups.google.com/forum/?utm_medium=email&utm_source=footer\#!msg/dedalus-users/Q63KhHSlfrI/fcFULv12AAAJ}} led to the development of a new tutorial problem\footnote{\url{https://bitbucket.org/dedalus-project/dedalus/pull-requests/44}}. This means that at least some of the time spent participating in answering questions on the mailing list is not a pure cost in terms of an additional burden on the core team--it can yield significant returns for the entire community.

\section{Software stack}
\label{sec:stack}

Complex AFD software relies on a non-trivial software stack to run. The stack can include compilers (e.g. C, C++, FORTRAN), interpreters (python), and libraries (FFTW, HDF5, BLAS, LAPACK). Besides providing input parameter files, analysis scripts, and the simulation code itself, one must also provide the entire software stack to outside researchers. In the case of \dedalus{}, the software stack is entirely open-source. This has both positive and negative effects on usability and workflow. It ensures there are no closed source or expensive, proprietary bottlenecks, and it ties to a rich ecosystem of tools, including Jupyter notebooks, docker containers, and integration with cloud computing services (e.g. \href{http://cocalc.com}{CoCalc} and \href{https://aws.amazon.com}{Amazon Web Services (AWS)}). 

However, a significant drawback is the complexity of \emph{installing} this stack. One of the greatest challenges we face is the distribution of software on heterogeneous computing environments, including laptops running macOS, university clusters from hundreds to thousands of cores, and national class supercomputing facilities in the United States (NASA, NSF, DoE) and Europe (PRACE). In addition to merely installing \dedalus{} on these systems, an additional challenge lies in optimizing for performance and verification testing to ensure accuracy. 

To support this effort, we have developed simple scripts to install both \dedalus{} and its stack. This approach is cumbersome to maintain and inefficient in both install time and disk space, and so we are pursuing several avenues to replace the scripts. A non-core team member has contributed a series of tools to allow \dedalus{} to integrate with the popular open-source \texttt{miniconda} package management system\footnote{\url{https://github.com/pkgw/dedalus-builder}}.

In order to allow users to assess optimization and performance, we have published test problems with scaling results and expected overall performance on key machines including the NASA Pleiades system\footnote{\url{https://bitbucket.org/exoweather/incompressible_ns_tg}}. 

\section{Looking to the Future}
\label{sec:future}

As mentioned in section~\ref{sec:repro}, there are often published results relying on calculations at a scale far too large to allow outside researchers to reproduce directly. In this case, we believe that the solution is to make datasets available along with analysis scripts and parameter files. The latter are particularly important should Moore's law continue to provide exponential growth in computational power; what is infeasible to reproduce today may not be in a few years' time. An open-code policy adapted by NASA should consider not only the requirement to make this available but also provide guidance on the use of \emph{middleware} designed to ease the burden of doing so. One particularly promising effort is the Whole Tale Project, which promises to link data, analysis scripts, and computing frontends (e.g. Jupyter notebook) to allow ``living publications'' in which any researcher can re-run analysis on datasets to reproduce results. 

Second, an open-code policy should also be focused on a future where computing hardware is increasingly fungible. Here, we imagine an emphasis on containerization solutions for software stacks including the open-source AFD simulation tools themselves. The use Docker containers for this purpose in industry is widespread, and astrophysics should follow suit. While some challenges remain for high-performance computing applications, \href{https://github.com/NERSC/shifter}{Shifter} and \href{http://singularity.lbl.gov/}{Singularity} have made significant steps toward remedying them. 

\section{Recommendations}
\label{sec:recommendations}
We conclude by enumerating our recommendations for an open code policy:
\begin{enumerate}
\item Require the \emph{explicit} use of an approved Open Source Initiative license (GPL, MIT, Apache, etc.)
\item Require the release of input files, analysis scripts, and build-time parameters to ensure reproducibility
\item Emphasize the importance of developing communities of practice around tools
\item Consider the future of containerized distribution and encourage the development of robust middleware to allow researchers to run analyses on remote data sources provided as part of publication.
\end{enumerate}

{\footnotesize
\bibliographystyle{aasjournal}
\bibliography{dedalus}}

\end{document}